\documentclass[final]{IEEEtran}

\usepackage{amssymb}
\usepackage{amsmath}
\usepackage{mathrsfs}
\usepackage{dsfont}

\DeclareMathOperator*{\argmax}{arg\,max}
\DeclareMathOperator*{\argmin}{arg\,min}
\DeclareMathOperator{\tr}{tr}
\DeclareMathOperator{\vectorize}{vec}
\DeclareMathOperator{\diag}{diag}

\newcommand{\mat}[1]{\boldsymbol{\mathrm{#1}}}

\newtheorem{thm}{Theorem}
\newtheorem{cor}{Corollary}
\newtheorem{lem}{Lemma}
\newtheorem{prop}{Proposition}
\newtheorem{remark}{Remark}
\newtheorem{defin}{Definition}
\newtheorem{example}{Example}
\begin{document}
\title{On the relaxed maximum-likelihood blind MIMO channel estimation for orthogonal space-time block codes}

\author{Kamran~Kalbasi and S.~Jamaloddin~Golestani}

\maketitle

\begin{abstract}
This paper concerns the maximum-likelihood channel estimation for MIMO systems with orthogonal space-time block codes when the finite alphabet constraint of the signal constellation is relaxed. We study the channel coefficients estimation subspace generated by this method. We provide an algebraic characterisation of this subspace which turns the optimization problem into a purely algebraic one and more importantly, leads to several interesting analytical proofs.  We prove that with probability one, the dimension of the estimation subspace for the channel coefficients is deterministic and it decreases by increasing the number of receive antennas up to a certain critical number of receive antennas, after which the dimension remains constant. In fact, we show that beyond this critical number of receive antennas, the estimation subspace for the channel coefficients is isometric to a fixed deterministic invariant space which can be easily computed for every specific OSTB code.
\end{abstract}
\begin{IEEEkeywords}
MIMO systems, orthogonal space-time block codes, blind channel estimation, relaxed maximum-likelihood, estimation subspace, Hurwitz-Radon family
\end{IEEEkeywords}

\section{Introduction}
Bandwidth limitation and channel fading are two major problems in wireless communication systems. In the late 90's, it was shown that the capacity of a fading channel is substantially increased by using multiple antennas in the transmitter and receiver \cite{Tse}. Such communication systems are referred to as multiple-input multiple-output (MIMO) systems. The gain in the capacity of a MIMO system can be attributed to the spatial diversity that such systems provide in coping with channel fading.  In order to approach the large theoretical capacity of a MIMO system, it is necessary to perform a combined form of coding and modulation involving both the time domain and the space domain (i.e. concerning multiple transmit antennas). This type of combined coding and modulation is referred to as space-time coding.

The most popular category of space-time codes is that of the orthogonal space-time block (OSTB) codes which have very rich algebraic structures. These codes, besides achieving high diversity gain, can be decoded by very simple symbol-by-symbol maximum-likelihood method.

A very important aspect of MIMO communication systems is the channel estimation without which correct decoding of the input data is impossible. In a MIMO system, channel estimation involves the estimation of channel coefficients corresponding to each pair of input-output antennas, all put together in a matrix which is called the channel matrix. The most popular way of obtaining this matrix is to send a training sequence, thus sacrificing a fraction of the transmission rate. On the other hand, due to rapid changes of the channel and/or limited resources, training and channel tracking may be infeasible. One possible remedy is to differentially encode the transmitted data and thus eliminate the need for channel knowledge. This latter method has a 3dB power loss, see e.g.  \cite{dif4}. Another way is to exploit known properties of the transmitted data to learn the channel blindly, which is called blind channel estimation, see e.g. \cite{SignalIdentification} and \cite{HigherOrderStatistics}.

One of the most natural strategies for blind channel estimation is through the maximum-likelihood method. The exact maximum-likelihood estimation which involves the maximization of the received signals' probability density taken over all possible input-data vectors and channel coefficients, has in general high computational complexity, especially for the OSTB codes with large blocks (see e.g. \cite{Ma2007} where a practical algorithm based on exact maximum likelihood has been proposed for a class of OSTB codes utilizing specific constellations). Another approach with much lower complexity is the relaxed maximum-likelihood which is similar to the exact maximum likelihood, but it relaxes the finite alphabet constraint (constellation) of the input signal. This method has been studied for the MIMO systems utilizing OSTB codes in \cite{CFE} where an equivalent formulation in terms of a matrix optimization was developed. It has also been used in other setups, see e.g. \cite{SarmadiPesaventoGershman2011}.

For the channel coefficients of a MIMO system with $N$ transmit antennas and $M$ receive antennas, we use interchangeably two different representations. The first is in the form of a complex matrix in $\mathbb{C}^{N\times M}$ which we call the channel matrix and denote by $\mat{H}_{\circ}$, and the second is in the form of a real-valued vector in $\mathbb{R}^{2MN}$ which we call the channel vector and denote by $\mat{h}_{\circ}$. There is a one-to-one correspondence between $\mat{h}_{\circ}$ and $\mat{H}_{\circ}$ via the linear isometry $\mat{h}_{\circ}=\underline{\mat{H}_{\circ}}$ to be defined in Section \ref{Linear Algebra}.
For any given OSTB code $\mathcal{C}$ and any channel matrix $\mat{H}_{\circ}\in \mathbb{R}^{M\times N}$ the relaxed maximum-likelihood method proposed in \cite{CFE} generates an estimation subspace for the channel coefficients (channel vector, to be precise) which we denote by $\boldsymbol{\mathcal{H}}^{\mathcal{C}}_{M}(\mat{H}_{\circ})$.  Clearly, the dimension of the space $\boldsymbol{\mathcal{H}}^{\mathcal{C}}_{M}(\mat{H}_{\circ})$ which naturally contains the true channel vector $\mat{h}_{\circ}$,  is of critical importance. For instance, if it happens to be equal to one, then the channel vector can be specified within a multiplicative constant, and this scalar factor can be further computed by other statistics \cite{CFE}.

As it has been demonstrated by numerical simulations in \cite{CFE}, the dimension of $\boldsymbol{\mathcal{H}}^{\mathcal{C}}_{M}(\mat{H}_{\circ})$ is not $1$ for some well-known OSTB codes including the Alamouti scheme.
So studying this estimation approach in the cases where this dimension is not $1$ can have interesting practical results as well as theoretical insights. In the current paper we try to shed some light on these cases.

To begin, we need to note that the space $\boldsymbol{\mathcal{H}}^{\mathcal{C}}_{M}(\mat{H}_{\circ})$
and its dimension depend on the realized channel vector $\mat{h}_{\circ}$, hence they are stochastic entities. Nevertheless the numerical simulations conducted in \cite{CFE} for many OSTB codes suggest that the dimension of this space interestingly shows a deterministic behavior for which no theoretical explanation has been provided. To our knowledge, the only other work that has studied this problem from a theoretical point of view is that of \cite{BI} where the deterministic behavior of the dimension of $\boldsymbol{\mathcal{H}}^{\mathcal{C}}_{M}(\mat{H}_{\circ})$ has been explained for a special class of OSTB codes which they call `identifiable from second-order statistics' \cite{BI}. In fact, these are precisely the OSTB codes for which the dimension of $\boldsymbol{\mathcal{H}}^{\mathcal{C}}_{M}(\mat{H}_{\circ})$ is $1$ for some channel matrix $\mat{H}_{\circ}$ and $M$ large enough.

From \cite{CFE}, we know that a vector $\mat{h}$ lies in $\boldsymbol{\mathcal{H}}^{\mathcal{C}}_{M}(\mat{H}_{\circ})$ if and only if $\mat{A}(\frac{\mat{h}}{\| \mat{h}\|})=\mat{A}(\frac{\mat{h}_{\circ}}{\| \mat{h}_{\circ}\|}) \mat{B}$ for some orthogonal matrix $\mat{B}$, where $\mat{A}(\cdot)$ is a linear map defined in Section \ref{Space-Time Block Coding}. Denoting the true input-data vector by $\mat{s}$, we can apply the algorithm in \cite{CFE} to find a vector $\hat{\mat{h}}$ in $\boldsymbol{\mathcal{H}}^{\mathcal{C}}_{M}(\mat{H}_{\circ})$ and use it as the channel vector to decode an estimated input-data vector $\hat{\mat{s}}$. Let $\hat{\mat{B}}$ be an orthogonal matrix that satisfies $\mat{A}(\frac{\hat{\mat{h}}}{\| \hat{\mat{h}}\|})=\mat{A}(\frac{\mat{h}_{\circ}}{\| \mat{h}_{\circ}\|}) \hat{\mat{B}}$. Then it is easy to show that we then have $\mat{s}=c\,\hat{\mat{B}}\,\hat{\mat{s}}$, where $c$ is some positive constant. This means that the true input-data vector $\mat{s}$ is an orthogonal transformation of $\hat{\mat{s}}$, up to a multiplicative constant. Can we say anything more about the set of all the possible matrices $\hat{\mat{B}}$? The main objective of this paper is to answer this question.

Indeed let us denote by
$\mathfrak{B}^{\mathcal{C}}_{M}( \mat{H}_{\circ})$ the space of all matrices $\mat{B}$ that are orthogonal up to a multiplicative constant and for which there exists a vector $\mat{h}_{\mat{B}}$ such that $\mat{A}(\mat{h}_{\mat{B}})=\mat{A}(\mat{h}_{\circ}) \mat{B}$. We show that $\mathfrak{B}^{\mathcal{C}}_{M}( \mat{H}_{\circ})$ is in fact isometric to $\boldsymbol{\mathcal{H}}^{\mathcal{C}}_{M}(\mat{H}_{\circ})$, hence sharing the same algebraic properties; for example they have the same dimension.

We derive a very simple algebraic characterisation for $\mathfrak{B}^{\mathcal{C}}_{M}( \mat{H}_{\circ})$. Indeed, we show that $$
\mathfrak{B}^{\mathcal{C}}_{M}(\mat{H}_{\circ})= \{\mat{B} \in \mathbb{R}^{K\times K}\,\big{|}\, \mat{\Gamma_B}\mat{H}_{\circ}=\mat{0}\},
$$
where $K$ is the size of input-data vectors, i.e. $\mat{s}\in \mathbb{R}^K$, and $\mat{\Gamma_B}$ is a matrix-valued linear function of $\mat{B}$ defined in Section \ref{Channel Estimation}.
As the spaces $\mathfrak{B}^{\mathcal{C}}_{M}( \mat{H}_{\circ})$ and $\boldsymbol{\mathcal{H}}^{\mathcal{C}}_{M}(\mat{H}_{\circ})$ are isometric, this representation provides us with a simple computable algebraic characterisation for $\boldsymbol{\mathcal{H}}^{\mathcal{C}}_{M}(\mat{H}_{\circ})$ as well. This turns our original optimization problem with intrinsic high complexity into a system of linear equations.

With the above characterisation of $\mathfrak{B}^{\mathcal{C}}_{M}( \mat{H}_{\circ})$ and using ideas from algebraic geometry, we show that with probability one, its dimension is a deterministic number. We also show that by increasing the number of receive antennas $M$, this dimension almost surely (i.e. with probability one) decreases until a certain critical number of receive antennas, which we denote by $M_*^{\mathcal{C}}$. These results are further validated by numerical simulations conducted for many OSTB codes, the outcome of which are tabulated in \cite[Table I]{CFE} and \cite[Table I]{BI}. We also note that these deterministic and almost sure properties have been proved in \cite{BI} only for the special case of codes that are \textit{identifiable from second-order statistics}, so our work generalizes their results to all OSTB codes.

Interestingly, with $M$ larger than or equal to $M_*^{\mathcal{C}}$, not only the  dimension of $\mathfrak{B}^{\mathcal{C}}_{M}( \mat{H}_{\circ})$ remains the same, but in fact the whole space $\mathfrak{B}^{\mathcal{C}}_{M}( \mat{H}_{\circ})$ becomes equal to an invariant deterministic space, denoted by $\mat{\mathscr{B}}^{\mathcal{C}}_*$ which is independent of the channel realisation. Again, it turns out that $\mat{\mathscr{B}}^{\mathcal{C}}_*$ itself has very simple structure. Indeed we show that
$$
\mat{\mathscr{B}}^{\mathcal{C}}_*= \{\mat{B} \in \mathbb{R}^{K\times K}\,\big{|}\, \mat{\Gamma_B}=\mat{0}\}.
$$
This enables us to explicitly compute the $\mat{\mathscr{B}}^{\mathcal{C}}_*$ for every specific OSTB code. As an example, we compute the space for the Alamouti scheme in Example \ref{Example}. Explicitly knowing the space $\mat{\mathscr{B}}^{\mathcal{C}}_*$ and by our isometry relation, the space $\boldsymbol{\mathcal{H}}^{\mathcal{C}}_{M}(\mat{H}_{\circ})$, could be potentially used in designing optimum training sequences for semi-blind MIMO systems, e.g. \cite{CompleteComplementarySequence}.

Our results could also have potential consequences for security issues studied in e.g. \cite{SecureSTBC} and \cite{TwoWayTraining}, because they suggest that when eavesdropping a communication link utilizing OSTB codes, one does not need to search the extensive set of all possible rotations, but just a relatively small linear space. For example, for the Alamouti code, the search space is a $4$-dimensional linear space.

We also show that the space $\mat{\mathscr{B}}^{\mathcal{C}}_*$ (and also $\mathfrak{B}^{\mathcal{C}}_{M}( \mat{H}_{\circ})$) has a basis consisting of the identity matric and the matrices of some Hurwitz-Radon family, a property that might be useful in decoding or even in designing constellation schemes that render the channel identifiable.

Finally, it should be noted that the class of OSTB codes that are \textit{identifiable from second-order statistics} introduced in \cite{BI} happens to be the same as the class of codes for which the dimension of $\mat{\mathscr{B}}^{\mathcal{C}}_*$ is one. This provides a very easy criterion to check the second-order-statistics identifiability of OSTB codes as defined in \cite{BI}.

\section{Preliminaries}
In this section, we gather some mathematical facts, notations, and definitions that we use in the sequel. We divide it into four subsections on linear algebra, optimization, algebraic geometry, and finally space-time block codes.
\subsection{Linear Algebra}\label{Linear Algebra}
We denote matrix \textit{transpose} by superscript $(\cdot)^\mathrm{T}$, \textit{conjugate transpose} by superscript $(\cdot)^\mathrm{H}$, the trace of a matrix by $\tr\{\cdot\}$, and finally the inverse of a matrix by $(\cdot)^{-1}$. The identity matrix of size $q$ is denoted by $\mat{I}_q$.\\
We also denote the space of all real matrices of size $m\times n$ by $\mathbb{R}^{m\times n}$ and the space of all complex matrices of size $m\times n$ by $\mathbb{C}^{m\times n}$. For complex numbers and matrices we use the notations $\Re(\cdot)$ and $\Im(\cdot)$ to denote their real and imaginary parts. A matrix $\mat{B}\in\mathbb{R}^{m\times m}$ is called \textit{orthogonal} if $\mathbf{B}^{\mathrm{-1}}=\mathbf{B}^{\mathrm{T}}$. We define the space $\mathscr{U}^{m\times q}$ as follows
$$
\mathscr{U}^{m\times q}:=\bigl\{\mat{Q}\in \mathbb{R}^{m\times q}\, \big{|}\, \mat{Q}^{\mathrm{T}}\mat{Q}=\mat{I}_q\bigr\}\,.
$$
Evidently, $\mathscr{U}^{n\times n}$ is the space of orthogonal matrices of size $n$.

We denote by $\vectorize(\cdot)$  the vectorization of a matrix, i.e. putting all the columns of a matrix (preserving the order) into a single column-vector. In other words, if $\mathbf{c}_1$, ..., $\mathbf{c}_n$ are the columns of matrix $\mathbf{M}$, we have $\vectorize(\mathbf{M}):=\left[\mathbf{c}_1^{\mathrm{T}}, \cdots, \mathbf{c}_n^{\mathrm{T}}\right]^{\mathrm{T}}$.

We define the \textit{underline} operator as follows: for any complex-valued matrix $\mat{P}$\
$$
\underline{\mat{P}}:= \vectorize\bigl(
                            \begin{bmatrix}
                            \Re(\mat{P})\\%[5pt]
                            \Im(\mat{P})
                            \end{bmatrix}\bigr).
$$

For any two arbitrary matrices $\mat{A}$ and $\mat{B}$, their Kronecker (or tensor) product \cite{AdabirMagnus,horn} denoted by $\otimes$ is defined by
$$
\mat{A}\otimes \mat{B}:=
\begin{bmatrix}
a_{11}\mat{B} & \cdots &a_{1m}\mat{B}\\
\vdots & \ddots &\vdots\\
a_{1n}\mat{B} & \cdots &a_{nm}\mat{B}\\
\end{bmatrix},
$$
where $a_{ij}$ denotes the $(i,j)$ entry of $\mat{A}$.

The \textit{overline} $\overline{ (\cdot) }$ operator is defined for any matrix $\mat{A} \in \mathbb{C}^{m\times n}$ as follows
$$
\begin{aligned}
\overline{\mat{A}} :=   \begin{bmatrix}
                            \Re(\mat{A}) & -\Im(\mat{A}) \\
                            \Im(\mat{A}) & \Re(\mat{A})
                            \end{bmatrix}
\end{aligned}.
$$

The following lemmas can be easily verified by the definitions given above; see e.g., \cite[ch. 10]{AdabirMagnus}.

\begin{lem}\label{algebraic lemma 1}
The underline operator is one-to-one and $\mathbb{R}$-linear. Moreover, for any two arbitrary complex-valued matrices $\mat{A}$ and $\mat{B}$ which are of the same size we have
$$\underline{\mat{A}}\,^{\mathrm{T}}\underline{\mat{B}}=
\tfrac{1}{2}\,tr\bigl\{\mat{A}^{\mathrm{H}}\mat{B}+\mat{B}^{\mathrm{H}}\mat{A}\bigr\}.
$$
\end{lem}

\begin{lem}\label{lemma3}
The overline operator is one-to-one and $\mathbb{R}$-linear. Moreover,
for any arbitrary complex-valued matrices $\mat{A}$ and $\mat{B}$, we have $\overline{\mat{A}^{\mathrm{H}}\,}={\overline{\mat{A}\,}}^{\mathrm{T}}$, and $\overline{\mat{A}\mat{B}}=\overline{\mat{A}}\,\, \overline{\mat{B}}$, provided that their sizes allow the multiplication.
\end{lem}

\begin{lem}\label{lemma4}
The Kronecker product is bilinear (i.e. linear in each of its arguments), and
for arbitrary matrices $\mat{A}$, $\mat{B}$, $\mat{C}$, and $\mat{D}$ the following equality holds (provided that the dimensions of the matrices allow the matrix multiplications)
\begin{equation}\label{multiplication of tensor products}
\bigl(\mat{A}\otimes \mat{B}\bigr) \bigl(\mat{C}\otimes\mat{D}\bigr)=\bigl(\mat{A}\mat{B}\bigl) \otimes \bigl(\mat{C}\mat{D}\bigr).
\end{equation}
\end{lem}

\begin{lem}\label{algebraic lemma 2}
For any two arbitrary complex or real-valued matrices $\mat{A}$ and $\mat{B}$ of sizes $m\times n$ and $n\times p$ respectively, we have
\addtocounter{equation}{1}
\begin{align}
\tag{\arabic{equation}a}\label{l1}\vectorize( \mat{A}\mat{B} )&=\bigl(\mat{B}^{\mathrm{T}} \otimes\mat{I}_m\bigr)\,\vectorize(\mat{A})\,,\\
\tag{\arabic{equation}b}\label{l2}\underline{\mat{A}\mat{B}}&=\bigl(\mat{I}_p \otimes\overline{\mat{A}}\bigr)\,\underline{\mat{B}}.
\end{align}
\end{lem}

\begin{lem}\label{lemma on kernels}
Let $\mathcal{X}$, $\mathcal{Y}$ and $\mathcal{Z}$ be arbitrary vector spaces, and $f: \mathcal{X} \rightarrow \mathcal{Y}$ and $g: \mathcal{Y} \rightarrow \mathcal{Z}$
be linear functions. Additionally assume that $\mathcal{X}$ has finite dimension. Then
$$
\dim \bigl(\ker (g\circ f)\bigr)=\dim \bigl(\ker (f)\bigr)+\dim \bigl(\ker (g|_{_{f[\mathcal{X}]}})\bigl)\,,
$$
where $\dim(\cdot)$ denotes the dimension of a space, $\ker(\cdot)$ the kernel(nullity) of a function, $f[\mathcal{X}]$ the image space of function $f$, and finally $g|_{_\mathcal{A}}$ denotes the restriction of function $g$ to subset $\mathcal{A}$.
\end{lem}
\begin{IEEEproof}
Let $\mathcal{K}:=\ker (g\circ f)$, and $f|_{_\mathcal{K}}: \mathcal{K} \rightarrow \ker (g|_{_{f[\mathcal{X}]}})$ be the restriction of $f$ to $\mathcal{K}$. The result follows from the well-known rank–nullity theorem (see e.g., \cite[Thm 2, ch. 3]{HoffmanKunze}).
\end{IEEEproof}

\subsection{Optimization}
The following theorem which goes back to Ky Fan \cite{Fan49}, plays a major role in studying the theoretical aspects of the relaxed maximum-likelihood channel estimation method introduced in \cite{CFE}. We need the following version which is slightly more general than the one stated in \cite{CFE}. As we were not able to find this version of the theorem or a satisfactory proof of it in the literature, we provide a proof in the appendix. Our proof relies only on elementary vector space ideas.

To begin, let $\mat{P}$ be a symmetric matrix in $\mathbb{R}^{m\times m}$, and $q$ be a
positive integer smaller than or equal to $m$; We define the space $\mathcal{A}_q^{\mat{P}}$ as follows
$$
\mathcal{A}_q^{\mat{P}}:=\argmax_{\mat{Q} \in \mathscr{U}^{m\times q}}\tr\bigl\{\mat{Q}^{\mathrm{T}}\mat{P}\mat{Q}\bigr\}\,,
$$
where  $\argmax$ denotes the set of arguments that maximize an expression.

Suppose that $\lambda_1$, $\lambda_2$, \dots, $\lambda_L$ list all the distinct eigenvalues of $\mat{P}$ in decreasing order, and for each $i$, $m_i$ be the multiplicity order of $\lambda_i$.
Let $k$ be the smallest positive integer satisfying
\begin{equation}\label{e6}
m_1+m_2+\dotsb+m_k\geq q\,,
\end{equation}
and let $\mathcal{V}_i$ be the eigenspace corresponding to $\lambda_i$, i.e.
$$
\mathcal{V}_i=\{\mat{v}\in \mathbb{R}^m\,\big{|}\,\mat{P}\mat{v}=\lambda_i\mat{v}\}\,. $$

With the above notations we have the following theorem.
\begin{thm}\label{thm1}
A matrix $\mat{Q} \in \mathscr{U}^{m\times q}$ is in $\mathcal{A}_q^{\mat{P}}$ if and only if
the vector space generated by all the columns of $\mat{Q}$ can be written as
$$
\mathcal{V}_1\oplus \dotsb \mathcal{V}_{k-1}\oplus \mathcal{W}_k\,,
$$
where $\mathcal{W}_k$ is a subspace of $\mathcal{V}_k$ of dimension ${q-(m_1+\dotsb+m_{k-1})}$, and $\oplus$ denotes the direct sum of vector spaces \cite{horn}.
\end{thm}
\begin{IEEEproof}
See appendix \ref{appen}.
\end{IEEEproof}
\begin{cor}\label{r1}
Suppose that for a given $\mat{P}$ in the above theorem,
the Inequality \eqref{e6} holds with equality, i.e. $m_1+m_2+\dotsb+m_k=q$ for some $k$, and let $\mat{Q}_\circ$ be an arbitrary matrix in $\mathcal{A}_q^{\mat{P}}$.
Then a matrix $\mat{Q}\in\mathscr{U}^{m\times q}$ is in $\mathcal{A}_q^{\mat{P}}$ if and only if there
exists an orthogonal matrix $\mathbf{B}\in \mathbb{R}^{q\times q}$  such that $\mat{Q}=\mat{Q}_\circ\mat{B}$.
\end{cor}
\begin{IEEEproof}
In this case $\mathcal{W}_k$ would be equal to $\mathcal{V}_k$, and so $\mat{Q}$
and $\mat{Q}_\circ$ have the same column space. As the matrices are both
in $\mathscr{U}^{m\times q}$ the result follows immediately.
\end{IEEEproof}

\subsection{Algebraic Geometry}
In this subsection we summarize some basic facts from algebraic geometry that we will need in the sequel. For more details we refer to \cite{AlgebraicGeometry}.

An \textit{algebraic set} is the locus of zeros of a finite collection of polynomials. In other words, any algebraic set is described as follows: let $\mathfrak{F}$ be a finite set of polynomials in $n$ variables with real coefficients. Then the following set is the algebraic set corresponding to $\mathfrak{F}$:
$$
\mathcal{S}_{\mathfrak{F}}:=\{\mathbf{x}\in\mathbb{R}^n\, \big{|}\,f(\mathbf{x})=0\,:\,\forall\, f\in \mathfrak{F}\}\,.
$$

Intuitively, every algebraic subset of $\mathbb{R}^n$ which is proper, i.e., it does not equal the whole space, is a hyper-surface of dimension at most $n-1$, and hence has no `volume'. The following theorem formalises this idea.
For measure theoretical concepts, we refer to e.g. \cite{Rudin}.
\begin{prop}\label{algebraic subsets have zero measure}
Every algebraic proper subset $\mathcal{S}\subsetneqq \mathbb{R}^n$ has zero Lebesgue measure in $\mathbb{R}^n$.
\end{prop}
\begin{IEEEproof}
It suffices to prove that for any non-zero polynomial $f$ of $n$ variables we have
$$
\mu_n(\{\mathbf{x}\in\mathbb{R}^n\, \big{|}\,f(\mathbf{x})=0\})=0\,,
$$
where $\mu_n$ denotes the Lebesgue measure on $\mathbb{R}^n$. We proceed by induction on $n$. For $n=1$ it is trivial. Now suppose the statement is true for every $k<n$. Then $f$ can be written as
$$
\begin{aligned}
f(x_n, x_{n-1}, \cdots, x_1)&=x_n^p \, g_p(x_{n-1}, \cdots, x_1)\\
&+x_{n}^{p-1} g_{p-1}(x_{n-1}, \cdots, x_1)\\
&+\cdots+g_0(x_{n-1}, \cdots, x_1)\,,
\end{aligned}
$$
where $g_i$'s are polynomials in the first $n-1$ variables. Let us denote $(x_{n-1}, \cdots, x_1)$ by $\mat{\bar{x}}_{n-1}$. Let also $\{f=0\}$, $\{g_p=0\}$, and
$\mathcal{S}_n$ be defined as follows
$$
\{f=0\}:=\{(x_n, \mat{\bar{x}}_{n-1}) \, \big{|}\, f(x_n, \mat{\bar{x}}_{n-1})=0\},
$$
$$
\{g_p=0\}:=\{\mat{\bar{x}}_{n-1} \, \big{|}\, g_p(\mat{\bar{x}}_{n-1})=0\},
$$
and
$$
\begin{aligned}
\mathcal{S}_n:=\{(x_n, \mat{\bar{x}}_{n-1}) \,\big{|}\, f(x_n, \mat{\bar{x}}_{n-1})=0\;
\text{and}\;
g_p(\mat{\bar{x}}_{n-1})\neq 0\}.
\end{aligned}
$$
Then we have
$$
\{f=0\}\subseteq \mathcal{S}_n \bigcup (\mathbb{R}\times \{g_p=0\}) \,.
$$

Notice that $\mu_n(\mathbb{R}\times \{g_p=0\})$ equals $\mu_{n-1}\{g_p=0\}$ which is zero by the induction hypothesis. Also by Fubini's theorem \cite{Rudin} we have
$$
\mu_n(\mathcal{S}_n)=\int_{\mathcal{A}_{n-1}}\mu_1\{x_n\big{|} f(x_{n},\mat{\bar{x}}_{n-1})=0\}\,\mathrm{d}\mu_{n-1}(\mat{\bar{x}}_{n-1})\,,
$$
where $\mathcal{A}_{n-1}:=\{(\mat{\bar{x}}_{n-1})\in\mathbb{R}^{n-1}\big{|}\,
g_p(\mat{\bar{x}}_{n-1})\neq 0\}$.
But for every $(\mat{\bar{x}}_{n-1})$ satisfying $g_p(\mat{\bar{x}}_{n-1})\neq 0$, the set $\{x_n\, \big{|}\, f(x_{n},\mat{\bar{x}}_{n-1})=0\}$ has at most $p$ elements, hence has zero Lebesgue measure.
\end{IEEEproof}
The above theorem justifies the following definition.
\begin{defin}
We call a subset of $\mathbb{R}^n$ algebraically negligible if it is contained in an algebraic proper subset of $\mathbb{R}^n$.
\end{defin}
\begin{remark}
The probability of any algebraically negligible subset of $\mathbb{R}^n$ is zero under any probability measure that is continuous with respect to the Lebesgue measure. In particular, any Gaussian probability on $\mathbb{R}^n$ with nonsingular correlation function falls into this category. Hence the probability of any algebraically negligible subset of $\mathbb{R}^n$ under a nonsingular Gaussian probability is zero.
\end{remark}

\subsection{Space-Time Block Coding}\label{Space-Time Block Coding}
We consider a MIMO system of ${N}$ transmit and ${M}$ receive antennas over a flat block-fading channel, i.e. the block length $L$ is much smaller than the channel coherence time. In this scenario, for the complex row-vectors $\mat{x}_t \in \mathbb{C}^{1 \times {N}}$, $\mat{y}_t\in \mathbb{C}^{1 \times M}$, and $\mat{\omega}_t\in \mathbb{C}^{1\times M}$ which represent respectively the transmitted, received, and noise signals at time slot $t$, $t=1, \cdots, L$, there exists an ${N}\times M$ complex matrix $\mat{H}_{\circ}\in \mathbb{C}^{{N}\times M}$ such that \cite{TO}
\begin{equation}\label{e1}
\mat{y}_t=\mat{x}_t\mat{H}_{\circ}+\mat{\omega}_t\,.
\end{equation}
The noise is assumed spatially and temporally white with a constant known variance $\sigma^2$ per complex dimension (or $\sigma^2/2$ per real dimension).

A space-time block code in real variables $\{s_i\}_{i=1}^K$ over ${N}$
transmit antennas of time block-length $L$, is a matrix-valued function $\mat{X}:\mathbb{R}^K\rightarrow \mathbb{C}^{L\times {N}}$ acting on input-data vectors $\mat{s}:=[s_1\,s_2\, \dots\, s_K]^{\mathrm{T}}\in \mathbb{R}^K$. Vector $\mat{s}$ represents the data to be encoded, and the $(i,j)$ element of $\mat{X}(\mat{s})$ represents the code symbol to be sent at time slot $i$ over transmit antenna $j$. It should be noted that this definition contains both the complex OSTB codes and the real ones in a unified manner, hence saving us from considering them separately.

Let $\mat{W}\in \mathbb{C}^{L\times {M}}$ represent the noise matrix at the receiver, i.e. its $(i,j)$ element represents the noise signal received at time slot $i$ in receive antenna $j$. Let also $\mat{Y}:\mathbb{R}^K\times \mathbb{C}^{L\times {M}}\rightarrow \mathbb{C}^{L\times {N}}$ denote the input-output function of the system, i.e. every $(i,j)$ element of $\mat{Y(s,W)}$ represents the received signal at time slot $i$ in receive antenna $j$ when the input-data vector is $\mat{s}$ and the noise matrix equals $\mat{W}$. With these notations \eqref{e1} can be written as
\begin{equation}\label{channel matrix equation}
\mat{Y(s,W)}=\mat{X(\mat{s})H}_{\circ}+\mat{W}\,.
\end{equation}
We should emphasize that as $\mat{H}_{\circ}$ is assumed to be constant over the transmission of the whole block, we do not explicitly denote it in $\mat{Y}$, although clearly $\mat{Y}$ it is a function of $\mat{H}_{\circ}$ as well.

A space-time block code $\mat{X}(\mat{s})$ is called orthogonal \cite{TO} if all the entries of $\mat{X}(\mat{s})$ are linear combinations of the variables $s_1,\;s_2\;\dots\;s_K$, and moreover for any arbitrary $\mat{s}\in \mathbb{R}^K$ the code matrix $\mat{X}(\mat{s})$ satisfies the following equation
\begin{equation}\label{e2}
\mat{X}(\mat{s})^{\mathrm{H}}\,\mat{X}(\mat{s})=\|\mat{s}\|^2\,\mat{I}_{N},
\end{equation}
where $\mat{I}_{N}$ is the identity matrix of size ${N}$.
In other words, in an orthogonal space-time block code the transmitted vectors on different transmit antennas are perpendicular to one another, and each has its norm equal to $\|\mat{s}\|$.\\
The linearity assumption is equivalent to
\begin{equation}\label{e3}
\mat{X}(\mat{s})=\sum^K_{k=1} s_k\,\mat{C}_k\,,
\end{equation}
where $\mat{C}_k \in \mathbb{C}^{L\times {N}}$ for $k=1, \cdots , K$.\\
Using the orthogonality property \eqref{e2}, we have
\begin{equation}\label{e4}
    \mat{C}_i^{\mathrm{H}}\,\mat{C}_j+ \mat{C}_j^{\mathrm{H}}\,\mat{C}_i=0; \quad \forall\,i,j \in \{1,\,\dots\,,K\}\,,\, i\neq j\,,
\end{equation}
and
\begin{equation}\label{e5}
\mat{C}_k^{\mathrm{H}}\,\mat{C}_k=\mat{I}_{N}; \quad \forall\, k=1,\,\dots\,,K.
\end{equation}

As the underline operator is $\mathbb{R}$-linear, by \eqref{channel matrix equation} and \eqref{e3} we have
\begin{equation}\label{MIMO system input-output relation}
\underline{\mat{Y(s,W)}}=\sum^K_{k=1} s_k \underline{\mat{C}_k\mat{H}_{\circ}}
+\underline{\mat{W}}.
\end{equation}

Let $\mat{H}\in \mathbb{C}^{{N}\times {M}}$ and define $\mat{h}:=\underline{\mat{H}}$. Using Equality \ref{l2}, we have
\begin{equation}\label{Definition of phi_k}
\underline{\mat{C}_k\mat{H}}=\mat{\Phi}_k\,\mat{h}\,,
\end{equation}
where $\mat{\Phi}_k$ is defined as follows
\begin{equation}\label{definition of Phi}
\mat{\Phi}_k:=
\mat{I}_{M}\otimes \overline{\mat{C}_k }\,.
\end{equation}

Using Equality \eqref{multiplication of tensor products} we obtain
\begin{equation}\label{e14}
\mat{\Phi}_i^{\mathrm{T}}\,\mat{\Phi}_j+
\mat{\Phi}_j^{\mathrm{T}}\,\mat{\Phi}_i=\mat{0};\quad \forall i,j\in\{1,2,\dots,K\}\,,\,i\neq j\,,
\end{equation}
and
\begin{equation}\label{e15}
\mat{\Phi}_k^{\mathrm{T}}\,\mat{\Phi}_k=\mat{I}_{2MN}; \quad \forall k\in\{1,2,\dots,K\}\,.
\end{equation}

Let us define
the matrix operator $\mat{A}:\mathbb{R}^{2NM}\rightarrow \mathbb{R}^{2TM \times K}$ as follows
\begin{equation}\label{Definition of A(h)}
\begin{aligned}
\mat{A(\mat{h})}:=
\begin{bmatrix}
\mat{\Phi}_1\mat{h}&\dots&\mat{\Phi}_K\mat{h}
\end{bmatrix}
;\quad \forall \mat{h}\in \mathbb{R}^{2NM}\,.
\end{aligned}
\end{equation}
With this definition, Equation \eqref{MIMO system input-output relation} can be written as follows
\begin{equation}\label{reveived vector formula}
\underline{\mat{Y(s,W)}}=\mat{A(\mat{h}_{\circ})}\,\mat{s}+\underline{\mat{W}},
\end{equation}
where $\mat{h}_\circ$ is related to $\mat{H}_\circ$ by $\mat{h}_\circ=\underline{\mat{H}_\circ}$.
Using Lemma \ref{algebraic lemma 1} and Equations \eqref{e14} and \eqref{e15}, we can easily show that for any $\mat{h}\in \mathbb{R}^{2NM}$ we have the following relation for the columns of $A(\mat{h})$
\begin{equation}\label{e7}
\begin{aligned}
(\mat{\Phi}_i\mat{h})^{\mathrm{T}}\mat{\Phi}_j\mat{h}&=
\tfrac{1}{2}tr\bigl\{\mat{h}^{\mathrm{T}}\bigl(\mat{\Phi}_i^{\mathrm{T}}
\mat{\Phi}_j+\mat{\Phi}_j^{\mathrm{T}}\mat{\Phi}_i\bigr)\mat{h}\bigr\}\\
&= \delta_{i,j}\,\| \mat{h}\|^2,
\end{aligned}
\end{equation}
where $\delta_{i,j}$ is the Kronecker delta.

As there is a direct one-to-one correspondence between $\mat{h}\in \mathbb{R}^{2NM}$ and $\mat{H}\in \mathbb{C}^{{N}\times {M}}$ via $\mat{h}=\underline{\mat{H}}$, we refer to $\mat{h}_{\circ}$ and $\mat{H}_{\circ}$ exchangeably as the channel coefficients, while calling $\mat{h}_{\circ}$ as the \textit{channel vector} and $\mat{H}_{\circ}$ as the \textit{channel matrix}.

\section{Channel Estimation}\label{Channel Estimation}
In this section we consider the blind channel estimation method first introduced for OSTB codes by Shahbazpanahi et al. \cite{CFE}. They found a closed-form estimation subspace which is in fact a formulation for the relaxed maximum likely estimation \cite{CFE,BI}. Indeed, suppose $\{\mat{y_i}\}_{i=1}^{J}$ denote $J$ consecutive blocks of received signals at the receiver, i.e., $\mat{y_i}:=\underline{\mat{Y}_i}$, and $\{\mat{s_i}\}_{i=1}^{J}$ denote the corresponding input-data vectors (to be estimated). Then by relaxing the constellation and finite alphabet constraint of the input-data vectors $\{\mat{s}_i\}_i$, one can show that the relaxed maximum likelihood estimation for $\mat{h}_{\circ}$ and $\{\mat{s_i}\}_{i=1}^{J}$ is equivalent to the following relaxed minimum-mean-square-error estimator \cite{CFE}
$$
\argmin_{\substack{\hat{\mat{h}_{\circ}}\in \mathbb{R}^{2MN} \\ \hat{\mat{s}}_1, \cdots, \hat{\mat{s}}_{J}\in \mathbb{R}^{K}}} \sum_{i=1}^{J} \|\mat{y}_i-\mat{A}(\hat{\mat{h}_{\circ}})\,\hat{\mat{s}}_i\|^2.
$$
Then one can show \cite{CFE} that it is equivalent for the estimated data vectors $\hat{\mat{s}}_i$ and the channel coefficients vector $\hat{\mat{h}}_{\circ}$ to satisfy the following equations
\begin{equation}\label{decoding the data signal}
\hat{\mat{s}}_i=\frac{\mat{A}(\hat{\mat{h}}_{\circ})^\mathrm{T} \mat{y}_i}{\|\hat{\mat{h}}_{\circ}\|^2}
\end{equation}
and
\begin{equation}\label{estimation of the channel vector}
\frac{\mat{A}(\hat{\mat{h}}_{\circ})}{\|\hat{\mat{h}}_{\circ}\|}\in
\argmax_{ \mat{h} \in \mathbb{R}^{2MN} } \tr\{\frac{\mat{A}(\mat{h})^{\mathrm{T}}}{ \|\mat{h}\|} \hat{\mat{R}} \frac{\mat{A}(\mat{h})}{ \|\mat{h}\|} \},
\end{equation}
where $\hat{\mat{R}}:=\frac{1}{J} \sum_{i=1}^{J}\mat{y}_i \, \mat{y}^{\mathrm{T}}_i$ is the sample variance of the received blocks.

For exploring the properties of this estimation scheme, one needs to replace  the sample variance $\hat{\mat{R}}$ with the ``theoretical'' one, i.e., $\mat{R}:=\mathbb{E}\bigl[\mat{y}\,\mat{y}^{\mathrm{T}}\bigr]
=\mathbb{E}\bigl[\underline{\mat{Y}}\,\underline{\mat{Y}}^{\mathrm{T}}\bigr]$ just as it has been done in \cite{CFE} and \cite{BI}. We start by defining the estimation subspace $\boldsymbol{\mathcal{H}}^{\mathcal{C}}_{M}(\mat{H}_{\circ})$ of an OSTB code $\mathcal{C}$ with $M$ receive antennas and the channel matrix $\mat{H}_{\circ}$ as follows
\begin{equation}\label{channel estimation equation}
\boldsymbol{\mathcal{H}}^{\mathcal{C}}_{M}(\mat{H}_{\circ}):= \argmax_{ \mat{h} \in \mathbb{R}^{2MN} } \tr\{\frac{\mat{A}(\mat{h})^{\mathrm{T}}}{ \|\mat{h}\|} \mat{R} \frac{\mat{A}(\mat{h})}{ \|\mat{h}\|} \}\,.
\end{equation}

We need the two important properties stated in Lemma \ref{eigen properties of matrix R} below, that were proved in \cite{CFE} under the assumption that the constellation of input-data vectors $\{\mat{s}_i\}_{i}$ is such that the components of $\mat{s}$ on each coordinate of $\mathbb{R}^K$ are statistically uncorrelated; in other words $\mathbb{E}\bigl[\mat{s}\,\mat{s}^{\mathrm{T}}\bigr]=\mat{\Lambda}_s$, where $\mat{\Lambda}_s$ is a diagonal matrix with strictly positive diagonals. But this constraint on the constellation seriously limits our choices, especially if we want to benefit from the correlation inside input-data vectors for the channel estimation. Fortunately, using the well-known fact that every symmetric real matrix is diagonalizable with an orthogonal matrix (see e.g. \cite{horn}), the proof can be modified to eliminate this constraint. Indeed, $\mathbb{E}\bigl[\mat{s}\,\mat{s}^{\mathrm{T}}\bigr]$ being symmetric, can be represented as $\mathbb{E}\bigl[\mat{s}\,\mat{s}^{\mathrm{T}}\bigr]
=\mat{U}\mat{\Lambda}_s\mat{U}^{\mathrm{T}}$, where $\mat{\Lambda}_s$ is a diagonal matrix with positive diagonals, and $\mat{U}\in \mathbb{R}^{K\times K}$ is an orthogonal matrix. Then the rest of the proof follows the same arguments as in \cite[Section B.]{CFE}.
\begin{lem}\label{eigen properties of matrix R}
Suppose $\mathbb{E}\bigl[\mat{s}\,\mat{s}^{\mathrm{T}}\bigr]$ is nonsingular, hence represented as $\mathbb{E}\bigl[\mat{s}\,\mat{s}^{\mathrm{T}}\bigr]
=\mat{U}\mat{\Lambda}_s\mat{U}^{\mathrm{T}}$, where $\mat{\Lambda}_s$ is a diagonal matrix with strictly positive diagonals, and $\mat{U}\in \mathbb{R}^{K\times K}$ is an orthogonal matrix. Then for any OSTB code $\mathcal{C}$ with any number of receive antennas $M$, and any channel matrix $\mat{H}_{\circ}$, we have\\
(i) The eigenvalues of the matrix $\mat{R}$ consist of all the diagonal elements of the matrix $\Lambda:= \|\mat{h}_{\circ}\|^2\Lambda_s
+\frac{\sigma^2}{2}\mat{I}_{K}$, as well as the value $\frac{\sigma^2}{2}$ with multiplicity order $2ML-K$.\\
(ii) For every $i=1,\cdots, K$, the $i$\textsuperscript{th} column of $\mat{A}(\mat{h}_{\circ})\mat{U}$ is an eigenvector of $\mat{R}$ with the $i$\textsuperscript{th} diagonal element of $\Lambda$ as its corresponding eigenvalue.
\end{lem}

Since the diagonal elements of $\Lambda$ are all strictly larger than $\frac{\sigma^2}{2}$, the matrix $\mat{R}$ falls into the assumptions of Corollary \ref{r1} with $m=2ML$ and $q=k=K$. So Lemma \ref{eigen properties of matrix R} and Corollary \ref{r1} imply that  $\frac{\mat{A}(\mat{h}_{\circ})}{\| \mat{h}_{\circ}\|}\in \mathcal{A}_K^{\mat{R}}$ and moreover, a matrix $\mathbf{Q}\in \mathscr{U}^{2ML\times K}$ lies in $\mathcal{A}_K^{\mat{R}}$ if and only if $\mathbf{Q}=\mat{A}(\frac{\mat{h}_{\circ}}{\| \mat{h}_{\circ}\|})\mat{B}$ for some orthogonal matrix $\mat{B}\in\mathbb{R}^{K\times K}$. Hence a fortiori, for any $\mat{h}\in \mathbb{R}^{2NM} $, we have $\mat{h} \in \boldsymbol{\mathcal{H}}^{\mathcal{C}}_{M}(\mat{H}_{\circ})$ if and only if $\mat{A}(\frac{\mat{h}}{\| \mat{h}\|})=\mat{A}(\frac{\mat{h}_{\circ}}{\| \mat{h}_{\circ}\|}) \mat{B}$ for some orthogonal matrix $\mat{B}\in\mathbb{R}^{K\times K}$.

Now suppose $\tilde{\mat{h}}\in\boldsymbol{\mathcal{H}}^{\mathcal{C}}_{M}(\mat{H}_{\circ})$. Then by the above argument, we can find an orthogonal matrix $\mat{B}_{\tilde{\mat{h}}}$ for which $\mat{A}(\frac{\tilde{\mat{h}}}{\| \tilde{\mat{h}}\|})=\mat{A}(\frac{\mat{h}_{\circ}}{\| \mat{h}_{\circ}\|})\mat{B}_{\tilde{\mat{h}}}$. On the other hand, by the estimation equation, Equation \eqref{decoding the data signal}, the estimated input-data vector $\tilde{\mat{s}}$ is $\frac{1}{\|\tilde{\mat{h}}\|^2} \mat{A}(\tilde{\mat{h}})^\mathrm{T} \mat{y}$. Also by Equation \eqref{reveived vector formula}, in the absence of noise we have $\mat{y}=\mat{A(\mat{h}_{\circ})}\,\mat{s}$, where $\mat{s}$ is the true input-data vector. So we have the following equality for the estimated input-data vector $\tilde{\mat{s}}$
\begin{equation}\label{rotated signal at the receiver}
\tilde{\mat{s}}=\frac{\|\mat{h}_{\circ}\|}{\|\tilde{\mat{h}}\|}
\mat{B}_{\tilde{\mat{h}}}^\mathrm{T} \,\mat{s}.
\end{equation}
So in the absence of noise, this channel estimation leads to an orthogonal transformation of the input-data vector. Therefore, any information about the set of all  $\mat{B}_{\tilde{\mat{h}}}$ with $\mat{\tilde{h}}\in\boldsymbol{\mathcal{H}}^{\mathcal{C}}_{M}(\mat{H}_{\circ})$ is of great interest. First we introduce a notation for this set: for any OSTB code $\mathcal{C}$, the number of receive antennas $M$ and channel matrix $\mat{H}_{\circ}$, we define $\mathfrak{B}^{\mathcal{C}}_{M}(\mat{H}_{\circ})$ as the set of all $\mat{B} \in \mathbb{R}^{K\times K}$ satisfying $\mat{B}^{\mathrm{T}}\mat{B}=c\mat{I}_K$ for some constant $c$ for which there exists a vector $\mat{h}_{\mat{B}}\in\mathbb{R}^{2MN}$ such that $\mat{A}(\mat{h}_{\mat{B}})=\mat{A}(\mat{h}_{\circ})\mat{B}$. In the sequel, we will explore the structure of this space. While doing this, we will also obtain an algebraic characterisation of $\boldsymbol{\mathcal{H}}^{\mathcal{C}}_{M}(\mat{H}_{\circ})$ which turns its variational nature into a purely algebraic problem. Using this characterisation, we will be able to prove several interesting properties of $\boldsymbol{\mathcal{H}}^{\mathcal{C}}_{M}(\mat{H}_{\circ})$.
We begin with the following technical lemma.

\begin{lem}\label{algebraic equivalent of A(h)=A(h)B}
For any two column vectors $\mat{\tilde{h}}, \mat{h}\in\mathbb{R}^{2MN}$ and any matrix $\mat{B}=[b_{i,j}]_{i,j}\in \mathbb{R}^{K\times K}$, we have
$$
\mat{A}(\mat{\tilde{h}})=\mat{A}(\mat{h})\mat{B}
$$
if and only if $\mat{\Gamma_B}\mat{H}=\mat{0}$ and
\begin{equation}\label{e8}
\mat{\tilde{h}}=\mat{\Phi}^{\mathrm{T}} \frac{\mat{B}^{\mathrm{T}}\otimes \mat{I}_{2ML}}{K}\mat{\Phi}\mat{h},
\end{equation}
where
$$
\mat{h}=\underline{\mat{H}}
\quad,\quad
\mat{\Phi}:=
\begin{bmatrix}
\mat{\Phi}_1 \\
\vdots \\
\mat{\Phi}_{K}
\end{bmatrix}
\quad,\quad
\mat{\Gamma_B}:=
\begin{bmatrix}
\mat{\Gamma}^1_{\mat{B}} \\
\vdots     \\
\mat{\Gamma}^K_{\mat{B}}
\end{bmatrix}\,,
$$
and
$$
\mat{\Gamma}^k_{\mat{B}} := \Bigl[ \frac{1}{K} \sum_{i=1}^{K} \sum_{j=1}^{K} b_{ji} \mat{C}_k \mat{C}_i^{\mathrm{H}} \mat{C}_j - \sum_{l=1}^{K} b_{lk}\mat{C}_l \Bigr].
$$
\end{lem}
\begin{IEEEproof}
First observe that the following are equivalent
\begin{equation}\label{e19}
\mat{A}(\mat{\tilde{h}})=\mat{A}(\mat{h})\mat{B},
\end{equation}
and
$$
\vectorize\bigl( \mat{A}(\mat{\tilde{h}})\bigr)=\vectorize\bigl(\mat{A}(\mat{h})\mat{B}\bigr).
$$
By \eqref{Definition of A(h)}, for any $\mat{\tilde{h}} \in \mathbb{R}^{2MN}$ we have:
$$
\vectorize\bigl( \mat{A}(\mat{\tilde{h}})\bigr)=\mat{\Phi}\,\mat{\tilde{h}}.
$$
So this along with Equality \eqref{l1} implies that \eqref{e19} is equivalent to
\begin{equation}\label{e9}
\begin{aligned}
\mat{\Phi}\mat{\tilde{h}}&=(\mat{B}^{\mathrm{T}}\otimes \mat{I}_{2ML}) \vectorize\bigl(\mat{A}(\mat{h}) \bigr) \\
                          &=(\mat{B}^{\mathrm{T}}\otimes \mat{I}_{2ML})\mat{\Phi}\mat{h}\,.
\end{aligned}
\end{equation}
By multiplying this Equation from the left by $\mat{\Phi}^{\mathrm{T}}$, and noting the following equation
$$
\mat{\Phi}^{\mathrm{T}}\mat{\Phi}=
\sum_{k=1}^{K}\mat{\Phi}_k^{\mathrm{T}}\mat{\Phi}_k=K\mat{I}_{2MN},
$$
we have
$$
\mat{\tilde{h}}=\mat{\Phi}^{\mathrm{T}} \frac{\mat{B}^{\mathrm{T}}\otimes \mat{I}_{2ML}}{K}\mat{\Phi}\mat{h}.
$$
Thus, we have shown that if for a vector $\mat{\tilde{h}}$, Equation \eqref{e19} is satisfied then it will inevitably have the form of \eqref{e8}.

Now we prove that for an arbitrary vector $\mat{h}$ and arbitrary matrix $\mat{B}$, if $\mat{\tilde{h}}$ is given by Equation \eqref{e8} then we have the following equivalence relation
$$
\mat{A}(\mat{\tilde{h}})=\mat{A}(\mat{h})\mat{B} \quad \Leftrightarrow \quad \mat{\Gamma_B}\mat{H}=\mat{0}.
$$
Indeed, by plugging $\mat{\tilde{h}}$ from \eqref{e8} into \eqref{e9} which is equivalent to \eqref{e19}, we find the following inequality
\begin{equation}\label{e10}
\mat{\Phi}\mat{\Phi}^{\mathrm{T}} \frac{\mat{B}^{\mathrm{T}}\otimes\mat{I}_{2ML}}{K}\mat{\Phi}\mat{h}
=(\mat{B}^{\mathrm{T}}\otimes\mat{I}_{2ML})\mat{\Phi}\mat{h}
\end{equation}
which is the same as
$$
\forall k \quad : \quad \mat{\Phi}_k\mat{\Phi}^{\mathrm{T}} \frac{\mat{B}^{\mathrm{T}}\otimes\mat{I}_{2ML}}{K}\mat{\Phi}\mat{h}
=\sum_{l=1}^{K}\,b_{lk}\mat{\Phi}_l\,\mat{h},
$$
%********************************************************************************************
where $b_{lk}$ is the $(l,k)$ entry of $\mat{B}$. On the other hand we have
$$
\mat{\Phi}^{\mathrm{T}} \frac{\mat{B}^{\mathrm{T}}\otimes \mat{I}_{2ML}}{K} \mat{\Phi}=
\frac{1}{K} \sum_i \sum_j b_{ji} \mat{\Phi}_i^{\mathrm{T}} \mat{\Phi}_j.
$$
So \eqref{e10} is equivalent to
\begin{equation}\label{e18}
\forall k \;:\; \Bigl[ \frac{1}{K} \sum_{i=1}^{K} \sum_{j=1}^{K} b_{ji} \mat{\Phi}_k \mat{\Phi}_i^{\mathrm{T}} \mat{\Phi}_j - \sum_{l=1}^{K} b_{lk}\mat{\Phi}_l \Bigr] \mat{h}=\mat{0}.
\end{equation}

Applying Lemmas \ref{lemma3} and \ref{lemma4}, one can easily verify that Equation \eqref{e18} is equivalent to
\begin{equation}\label{e11}
\forall k \quad : \quad (\mat{I}_{M}\otimes \overline{\mat{\Gamma}^k_{\mat{B}}})\,
\underline{\mat{H}}=\mat{0}.
\end{equation}
By Equality \eqref{l2}, this is the same as
$$
\forall k=1,\dots,K \quad : \quad \underline{\mat{\Gamma}^k_{\mat{B}} \mat{H}}=\mat{0},
$$
or equivalently
$$
\mat{\Gamma_B}\mat{H}=\mat{0}.
$$
\end{IEEEproof}

\begin{remark}
In the course of the above proof one can easily verify the following representation for $\mat{\Gamma_B}$
$$
\mat{\Gamma_B}=\Bigl(\frac{1}{K}
\begin{bmatrix}
\mat{C}_1 \\
\vdots     \\
\mat{C}_K
\end{bmatrix}
\begin{bmatrix}
\mat{C}_1^\mathrm{H} & \hdots & \mat{C}_K^\mathrm{H}
\end{bmatrix}
-\mat{I}_{LK}
\Bigr) \, \mat{B}\mathrm{^T}\otimes \mat{I}_{L}
\begin{bmatrix}
\mat{C}_1 \\
\vdots     \\
\mat{C}_K
\end{bmatrix}.
$$
\end{remark}

\begin{prop}\label{prop1}
A column vector $\mat{h}\in\mathbb{R}^{2MN}$ is in $\boldsymbol{\mathcal{H}}^{\mathcal{C}}_{M}(\mat{H}_{\circ})$  if and only if there exists an orthogonal matrix $\mat{B}\in \mathbb{R}^{K\times K}$ such that
$$
\mat{\Gamma_B}\mat{H}_{\circ}=\mat{0}\quad \text{and} \quad
\frac{\mat{h}}{\|\mat{h}\|}=\mat{\Phi}^{\mathrm{T}} \frac{\mat{B}^{\mathrm{T}}\otimes \mat{I}_{2ML}}{K}\mat{\Phi}\frac{\mat{h}_{\circ}}{\|\mat{h}_{\circ}\|} .
$$
\end{prop}
\begin{IEEEproof}
Since $\frac{\mat{A}(\mat{h}_{\circ})}{\|\mat{h}_{\circ}\|}$ lies in $\mathcal{A}_q^{\mat{R}}$, a vector $\mat{h}\in\mathbb{R}^{2MN}$
is in $\boldsymbol{\mathcal{H}}^{\mathcal{C}}_{M}(\mat{H}_{\circ})$ if and only if $\frac{\mat{A}(\mat{h)}}{\|\mat{h}\|}$
lies in $\mathcal{A}_q^{\mat{R}}$.
But as we mentioned earlier, matrix $\mat{R}$ has exactly $K$ eigenvalues greater than $\frac{\sigma^2}{2}$ and ($2ML-K$) eigenvalues equal to $\frac{\sigma^2}{2}$; hence Corollary \ref{r1} applies. This implies that $\mat{h}$ is in $\boldsymbol{\mathcal{H}}^{\mathcal{C}}_{M}(\mat{H}_{\circ})$ if and only if there exists an orthogonal matrix $\mat{B}\in \mathbb{R}^{K\times K}$ such that
\begin{equation}\label{e13}
\frac{\mat{A}(\mat{h)}}{ \|\mat{h}\|}=\frac{\mat{A}(\mat{h}_{\circ})}{ \|\mat{h}_{\circ}\|} \mat{B}.
\end{equation}
By linearity of $\mat{A}(\cdot)$ we have
\begin{equation}\label{e20}
\mat{A}\bigl(\frac{\mat{h}}{ \|\mat{h}\|}\bigr)=\mat{A}\bigl(\frac{\mat{h}_{\circ}}{ \|\mat{h}_{\circ}\|}\bigr) \mat{B}.
\end{equation}
Now the result follows from Lemma \ref{algebraic equivalent of A(h)=A(h)B}.
\end{IEEEproof}
%*******************************************************************************************
As a consequence, we have the following interesting theorem
\begin{thm}
For any OSTB code $\mathcal{C}$, any number of receive antennas $M$, and any realized channel matrix $\mat{H}_{\circ}$, we have
$$
\mathfrak{B}^{\mathcal{C}}_{M}(\mat{H}_{\circ})= \{\mat{B} \in \mathbb{R}^{K\times K}\,\big{|}\, \mat{\Gamma_B}\mat{H}_{\circ}=\mat{0}\}.
$$
\end{thm}

For any OSTB code $\mathcal{C}$ and any realized channel matrix $\mat{H}_{\circ}$, we also define
$$
\mat{\mathscr{B}}^{\mathcal{C}}_* := \{\mat{B}\in \mathbb{R}^{K\times K} \,\big{|}\, \mat{\Gamma_B}=\mat{0} \}.
$$
It is evident that for any arbitrary code $\mathcal{C}$ and arbitrary channel realization $\mat{H}_{\circ}$ we have:
\begin{equation}\label{e17}
\mat{\mathscr{B}}^{\mathcal{C}}_* \subseteq \mathfrak{B}^{\mathcal{C}}_{M}( \mat{H}_{\circ}).
\end{equation}

The following theorem establishes an isometry, hence a one-to-one correspondence between $\mathfrak{B}^{\mathcal{C}}_{M}( \mat{H}_{\circ})$ and the space $\boldsymbol{\mathcal{H}}^{\mathcal{C}}_{M}(\mat{H}_{\circ})$.
\begin{thm}\label{isometry}
For any arbitrary OSTB code $\mathcal{C}$ and any channel realization $\mat{H}_{\circ}$, the map
$$\mat{B}\mapsto \mat{h}:=\mat{\Phi}^{\mathrm{T}} \frac{\mat{B}^{\mathrm{T}}\otimes \mat{I}_{2ML}}{K}\mat{\Phi}\mat{h}_{\circ}$$
is an isometry from $\mathfrak{B}^{\mathcal{C}}_{M}( \mat{H}_{\circ})$ onto $\boldsymbol{\mathcal{H}}^{\mathcal{C}}_{M}(\mat{H}_{\circ})$. As a result these spaces have the same dimension.
\end{thm}
\begin{IEEEproof}
First we show that every nonzero element of $\mathfrak{B}^{\mathcal{C}}_{M}( \mat{H}_{\circ})$ is in fact an orthogonal matrix up to a positive multiplicative constant. For any
arbitrary $\mat{B}$ in $\mathfrak{B}^{\mathcal{C}}_{M}( \mat{H}_{\circ})$ take its corresponding $\mat{h}$ given by Equation \eqref{e8} (with $\mat{h}=\underline{\mat{H}}$, as before). So by Lemma \ref{algebraic equivalent of A(h)=A(h)B}, we have
$$
\mat{A}(\mat{h})=\mat{A}(\mat{h}_{\circ}) \mat{B}.
$$
Hence we get
$$
\mat{A(h)}^{\mathrm{T}}\mat{A(h)}=\mat{B}^{\mathrm{T}} \mat{A(h_{\circ})}^{\mathrm{T}} \mat{A(h_{\circ})} \mat{B}.
$$
But by \eqref{Definition of A(h)} and \eqref{e7}, for any arbitrary vector $\mat{h}\in\mathbb{R}^{2MN}$ we have
$$
\mat{A(h)}^{\mathrm{T}}\mat{A(h)}=\|\mat{h}\|^2 \mat{I}_{K}.
$$
So we get
$$
\mat{B}^{\mathrm{T}}\mat{B}=\frac{ \|\mat{h}\|^2 }{ \|\mat{h}_{\circ}\|^2 } \, \mat{I}_{K},
$$
which shows that $\mat{B}$ is orthogonal up to a positive constant. So Proposition \ref{isometry} applies with $\frac{\mat{B}}{\|\mat{B}\|}$, and $\mat{h}$ lies in $\boldsymbol{\mathcal{H}}^{\mathcal{C}}_{M}(\mat{H}_{\circ})$. This correspondence is in fact an isometry between $\mathfrak{B}^{\mathcal{C}}_{M}( \mat{H}_{\circ})$ and $\boldsymbol{\mathcal{H}}^{\mathcal{C}}_{M}(\mat{H}_{\circ})$. Indeed, let us define an inner product on $\mathfrak{B}^{\mathcal{C}}_{M}( \mat{H}_{\circ})$ as follows:
For any $\mat{B}_1, \mat{B}_2\in \mathfrak{B}^{\mathcal{C}}_{M}( \mat{H}_{\circ})$
$$
\langle \mat{B}_1, \mat{B}_2\rangle:=\frac{\|\mat{h}_{\circ}\|^2}{K} tr\{\mat{B}_1^{\mathrm{T}} \mat{B}_2\}.
$$
Pursuant to \eqref{e8}, we have
$$
i=1,2 \quad: \quad \mat{h}_i := \mat{\Phi}^{\mathrm{T}} \frac{\mat{B}_i^{\mathrm{T}}\otimes \mat{I}_{2ML}}{K}\mat{\Phi}\mat{h}_{\circ}.
$$
Then we have
$$
\mat{h}_1^{\mathrm{T}} \mat{h}_2 =  \mat{h}_{\circ}^{\mathrm{T}} \mat{\Phi}^{\mathrm{T}} \frac{\mat{B}_1\otimes \mat{I}_{2ML}}{K}\mat{\Phi}   \mat{\Phi}^{\mathrm{T}} \frac{\mat{B}_2^{\mathrm{T}}\otimes \mat{I}_{2ML}}{K}\mat{\Phi}\mat{h}_{\circ}.
$$
By \eqref{e10} and \eqref{multiplication of tensor products}, the right-hand-side of this equation is equal to
\begin{equation}\label{e16}
\mat{h}_{\circ}^{\mathrm{T}} \mat{\Phi}^{\mathrm{T}} \frac{(\mat{B}_1\mat{B}_2^{\mathrm{T}})\otimes \mat{I}_{2ML}}{K} \mat{\Phi}\mat{h}_{\circ}.
\end{equation}
By expanding this expression and using Equality \eqref{e15} we have
$$
\mat{\Phi}^{\mathrm{T}} \frac{(\mat{B}_1\mat{B}_2^{\mathrm{T}})\otimes \mat{I}_{2ML}}{K} \mat{\Phi}=
\frac{ tr\{\mat{B}_1^{\mathrm{T}}\mat{B}_2\} }{K} \mat{I}_{2MN}+\mat{S},
$$
where
$$
\mat{S} := \frac{1}{K} \sum_{i \neq j}\; \beta_{i,j}\; \mat{\Phi}_i^{\mathrm{T}} \mat{\Phi}_j.
$$
Here $\beta_{i,j}$ denotes the $(i,j)$ entry of $\mat{B}_1^{\mathrm{T}} \mat{B}_2$.

By Equation \eqref{e14}, $\mat{S}$ is skew-symmetric, i.e.,
$$
\mat{S}^{\mathrm{T}}=-\mat{S}.
$$
So
$$
\mat{h}_{\circ}^{\mathrm{T}} \mat{S} \mat{h}_{\circ} = 0.
$$
Therefore, Equation \eqref{e16} reduces to
$$
\mat{h}_1^{\mathrm{T}} \mat{h}_2 = \frac{ tr\{\mat{B}_1^{\mathrm{T}}\mat{B}_2\} }{K} \|\mat{h}_{\circ}\|^2=\langle\mat{B}_1, \mat{B}_2\rangle.
$$
\end{IEEEproof}

We continue to explore the properties of these spaces.
First, notice that when ${M}\geq {N}$, the matrix $\mat{H}_{\circ}$ has with probability one ${N}$ linearly independent columns. The reason is that the columns of $\mat{H}_{\circ}$ are assumed to be jointly normal and stochastically independent. So in this case, we have the following implication
$$
\mat{\Gamma_B}\mat{H}_{\circ}=\mat{0} \quad \Longrightarrow \quad  \mat{\Gamma_B}=\mat{0} \quad \text{(with probability one)}.
$$
This shows that when ${M}\geq {N}$, we have
$$
\mathfrak{B}^{\mathcal{C}}_{M}( \mat{H}_{\circ}) = \mat{\mathscr{B}}^{\mathcal{C}}_* \quad \text{(with probability one)}.
$$
As $\mat{\mathscr{B}}^{\mathcal{C}}_*$ is according to its definition, independent of the channel matrix or even the number of receive antennas, this equality verifies that as far as ${M}\geq {N}$, the dimension of the estimation subspace is invariant of the number of receive antennas. This motivates the following definition.
\begin{defin}\label{Definition of M_*}
For an OSTB code $\mathcal{C}$, we denote the dimension of $\mat{\mathscr{B}}^{\mathcal{C}}_*$ by $d_*^\mathcal{C}$, and denote by $M_*^\mathcal{C}$ the smallest number of receive antennas for which the space $\mathfrak{B}^{\mathcal{C}}_{M_*^\mathcal{C}}( \mat{H}_{\circ})$ is equal to $\mat{\mathscr{B}}^{\mathcal{C}}_*$.
\end{defin}

Clearly, the dimension of $\boldsymbol{\mathcal{H}}^{\mathcal{C}}_{M}(\mat{H}_{\circ})$ and $M_*^\mathcal{C}$ are both random numbers. Nevertheless, the following theorem paves the way to prove that $\dim(\boldsymbol{\mathcal{H}}^{\mathcal{C}}_{M}(\mat{H}_{\circ}))$ and hence $M_*^\mathcal{C}$ are almost surely (i.e. with probability one) deterministic.

\begin{thm}\label{Kernel is deterministic}
Let $\mathcal{Y}$ be a finite-dimensional vector space, and $g:\mathbb{R}^{n}\times\mathcal{Y}\rightarrow\mathbb{R}^p$ be a bilinear map. For every $\mat{x}\in\mathbb{R}^{n}$, define the function $g_{\mat{x}}:\mathcal{Y}\rightarrow\mathbb{R}^p$ as $g_{\mat{x}}(\mat{y}):=g(\mat{x},\mat{y})$. Then we have\\
(i) There exists a non-negative integer $d_{g}$ such that the dimension of the kernel of $g_{\mat{x}}$ equals $d_{g}$ for every $\mat{x}\in \mathbb{R}^n$ except for an algebraically negligible subset of $\mathbb{R}^n$.\\
(ii) For every nontrivial $g$ (i.e. $g(\mat{x},\mat{y})\neq\mat{0}$ for some $\mat{x}$ and $\mat{y}$), $d_{g}$ is strictly smaller than $\dim(\mathcal{Y})$.
\end{thm}
\begin{IEEEproof}
Let $m$ be the dimension of $\mathcal{Y}$, and $\{\mat{y}_i\}_{i=1}^{m}$ be a basis for $\mathcal{Y}$. As $g_{\mat{x}}$ is linear for every $\mat{x}$, by the rank–nullity theorem \cite{HoffmanKunze} we have $\dim\bigl(\ker(g_{\mat{x}})\bigr)=m-\dim\bigl(g_{\mat{x}}[\mathcal{Y}]\bigr)$, where $g_{\mat{x}}[\mathcal{Y}]$ is the image space of $g_{\mat{x}}$. So it suffices to prove that the dimension of $g_{\mat{x}}[\mathcal{Y}]$ is constant for every $\mat{x}\in\mathbb{R}^n$ except for an algebraically negligible subset of $\mathbb{R}^n$. But the dimension of $g_{\mat{x}}[\mathcal{Y}]$ equals the rank of the matrix $\mat{\Xi}_{\mat{x}}:=\bigl[g(\mat{x},\mat{y}_1) \cdots g(\mat{x},\mat{y}_m)\bigr]$, i.e., the matrix composed of columns vectors $\{g(\mat{x},\mat{y}_i)\}_{i=1}^{m}$. We denote by $\mathfrak{S}_{\mat{\Xi}_{\mat{x}},k}$ the set of all sub-matrices of matrix $\mat{\Xi}_{\mat{x}}$ of size $k\times k$.  We now define $\mathcal{X}_i$ as follows
$$
\mathcal{X}_k:=\{\mat{x}\in\mathbb{R}^n\;\big{|}\; \det(\mat{T})=0:\; \forall\, \mat{T}\in \mathfrak{S}_{\mat{\Xi}_{\mat{x}},k}\},
$$
with the additional convention $\mathcal{X}_0:=\mat{0}$. Clearly this is an increasing sequence, i.e. $\mathcal{X}_{k}\subseteq \mathcal{X}_{k+1}$ for every $k$, and stabilizes to $\mathbb{R}^n$ at some point. We define $d_{g}$ as the largest integer for which  $\mathcal{X}_{d_{g}}\neq \mathbb{R}^n$. We also know that the rank of a matrix is the largest integer $k$ such that the matrix contains a non-singular sub-matrix of size $k\times k$ (see e.g. \cite{horn}). So for each $k$, the subset $\mathcal{X}_{k+1}\setminus \mathcal{X}_{k}$ is precisely the set of all $\mat{x}$ for which the rank of $\mat{\Xi}_{\mat{x}}$ equals $k$. On the other hand, each $\mathcal{X}_k$ is an algebraic subset of $\mathbb{R}^m$ because the determinant of a matrix is a polynomial function of its entries. Hence for each $\mathcal{X}_k$, if it is a proper subset of $\mathbb{R}^n$, it is inevitably algebraically negligible. In particular, as $\mathcal{X}_{d_{g}}$ is not the whole space by definition, it is algebraically negligible. This means that the rank of $\mat{\Xi}_{\mat{x}}$ equals $d_{g}$ for every $\mat{x}\in\mathbb{R}^n$ except for an algebraically negligible set.\\
To prove the second part, we should show that $d_{g}\neq m$. Let $\mathcal{X}_-$ be the set of all $\mat{x}\in\mathbb{R}^n$ such that $\dim\bigl(\ker (g_{\mat{x}})\bigr)=m$, or equivalently $\ker (g_{\mat{x}})=\mathcal{Y}$. As $g$ is non-trivial, there exists $\mat{y}_\circ\in\mathcal{Y}$ and $\mat{x}_\circ\in\mathbb{R}^n$ such that $g(\mat{x}_\circ,\mat{y}_\circ)\neq\mat{0}$. So in particular $\mathcal{X}_-\subseteq \mathcal{X}_{\mat{y}_\circ}:=\{\mat{x}\in\mathbb{R}^n;\quad g(\mat{x},\mat{y}_\circ)=\mat{0}\}$. Again as $\mathcal{X}_{\mat{y}_\circ}$ is an algebraic subset of $\mathbb{R}^n$ which is not the whole space, it is algebraically negligible.
\end{IEEEproof}

\begin{thm}
For any code $\mathcal{C}$ and any number of receive antennas $M$, the dimension of  $\boldsymbol{\mathcal{H}}^{\mathcal{C}}_{M}(\mat{H}_{\circ})$ equals a deterministic number (denoted by $\mathrm{d}^{\mathcal{C}}({M})$), almost surely for every $\mat{H}_{\circ}\in\mathbb{C}^{{N}\times M}$. Moreover, $\mathrm{d}^{\mathcal{C}}(M)$ is a decreasing function in ${M}\in \mathbb{N}$ as long as $M<M_*^\mathcal{C}$.
\end{thm}
\begin{IEEEproof}
By Theorem \ref{isometry}, the dimension of $\boldsymbol{\mathcal{H}}^{\mathcal{C}}_{M}(\mat{H}_{\circ})$ is the same as the dimension of $\mathfrak{B}^{\mathcal{C}}_M( \mat{H}_{\circ})$.
By Equality \eqref{l2}, for any $\mat{B}\in \mathbb{R}^{K\times K}$, the equality $\mat{\Gamma_B}\mat{H}_{\circ}=\mat{0}$ holds if and only if $(\mat{I}_{M} \otimes \overline{\mat{\Gamma}}_{\mat{B}})\,\underline{\mat{H}_\circ}=\mat{0}$. So by Lemma \ref{lemma on kernels}, we have
$$
\dim\bigl(\mathfrak{B}^{\mathcal{C}}_M( \mat{H}_{\circ})\bigr)=\dim\bigl(\mat{\mathscr{B}}^{\mathcal{C}}_*\bigr)+
\dim\bigl(\ker(g_{_{\mat{H}_\circ}})\bigr),
$$
where $g_{_{\mat{H}_\circ}}^M$ is the function $g_{_{\mat{H}_\circ}}:f[\mathbb{R}^{K\times K}]\rightarrow \mathbb{R}^{2MN}$ defined by $g_{_{\mat{H}_\circ}}(\mat{G})=(\mat{I}_{M} \otimes\mat{G})\underline{\mat{H}_\circ}$ for every $\mat{G}\in f[\mathbb{R}^{K\times K}]$, i.e., the image of
$f:\mathbb{R}^{K\times K}\rightarrow \mathbb{R}^{2LK\times 2N}$ defined by $f(\mat{B}):= \overline{\mat{\Gamma}}_{\mat{B}}$ for every $\mat{B}\in \mathbb{R}^{K\times K}$.

By Theorem \ref{Kernel is deterministic}, it is clear that there is a non-negative integer $\mathrm{d}_{g}^M$ such that the dimension of $\ker(g_{_{\mat{H}_\circ}})$ is equal to $\mathrm{d}_{g}^M$ for every $\underline{\mat{H}_\circ}\in\mathbb{R}^{2MN}$ except for an algebraically negligible subset. As $\mat{\mathscr{B}}^{\mathcal{C}}_*$ is independent of the channel coefficients or even $M$, this implies the existence of the deterministic dimension $\mathrm{d}^{\mathcal{C}}(M):=d_*^\mathcal{C}+\mathrm{d}_{g}^M$ for $\mathfrak{B}^{\mathcal{C}}_M( \mat{H}_{\circ})$ as described above.\\
For the second part, we should show that as long as it is strictly positive, $\mathrm{d}_{g}^M$ decreases by increasing $M$.

For every $\mat{H}_\circ\in\mathbb{C}^{M\times N}$ and $\mat{h}\in\mathbb{C}^{M}$, define $\mat{H}^+:=\big[\mat{H}_{\circ}\; \mat{h} \bigr]$, i.e. the channel matrix corresponding to $M+1$ receive antennas. Let $g_{_{\mat{H}^+}}$ and $g_{_{\mat{h}}}$ be defined in the same manner as above. Then we can easily verify that
$$
\ker(g_{_{\mat{H}^+}})=\ker(g_{\mat{h}}^{\mat{H}_\circ}),
$$
where $g_{\mat{h}}^{\mat{H}_\circ}$ is the restriction of $g_{_{\mat{h}}}$ to $\ker(g_{\mat{H}_\circ})$. By the second part of Theorem \ref{Kernel is deterministic}, as long as $\ker(g_{\mat{H}_\circ})\neq \{\mat{0}\}$, the dimension of $\ker(g_{\mat{h}}^{\mat{H}_\circ})$ is strictly smaller than that of $\ker(g_{\mat{H}_\circ})$.
\end{IEEEproof}

The next theorem shows that $\mathfrak{B}^{\mathcal{C}}_{M}( \mat{H}_{\circ})$ and $\mat{\mathscr{B}}^{\mathcal{C}}_*$ have very special algebraic structures. We remind the reader that a family of $n\times n$ real matrices $\mat{A}_1, \cdots, \mat{A}_k$ is called a Hurwitz-Radon family \cite{TO}, if $\mat{A}_i^\mathrm{T}=-\mat{A}_i$, $\mat{A}_i^2=-\mat{I}$ and $\mat{A}_i\mat{A}_j+\mat{A}_j\mat{A}_i=\mat{0}$ for every $i\neq j$. By Hurwitz-Radon theorem, any Hurwitz-Radon family of $n\times n$ matrices contains at most $\rho(n)-1$ matrices \cite{TO}, where $\rho(n)$ is the Hurwitz-Radon function defined as follows: for a positive integer $n$, if $n=2^a b$, b odd, and $a=4d+c$ where $a$, $b$, $c$, $d$ are non-negative integers with $0\leq c<4$, then $\rho(n)=2^c+8d$. So we are ready for the following theorem.

\begin{thm}\label{Hurwitz-Radon Basis and Pure Rotation}
The spaces $\mathfrak{B}^{\mathcal{C}}_{M}( \mat{H}_{\circ})$ and $\mat{\mathscr{B}}^{\mathcal{C}}_*$, each have a basis consisting of the identity matrix $\mat{I}_K$ and the matrices of a Hurwitz-Radon family. Moreover, when $K$ is even, any non-zero matrix $\mat{B}\in \mathbb{R}^{K\times K}$ in any of these spaces is a pure rotation up to a positive multiplicative constant, i.e., $\mat{B}\mathrm{^T}\mat{B}=c \mat{I}_{K}$ for some $c>0$, and $\det(\mat{B})>0$.
\end{thm}
\begin{IEEEproof}
As shown in the proof of Theorem \ref{isometry}, each element of $\mathfrak{B}^{\mathcal{C}}_{M}( \mat{H}_{\circ})$ is orthogonal up to a positive multiplicative constant. Consider the following procedure to construct a basis for $\mathfrak{B}^{\mathcal{C}}_{M}( \mat{H}_{\circ})$. The first element is trivially the identity matrix $\mat{I}_K$, which we denote by $\mat{B}_0$. For the next element, let $\mat{B}_1$ be an arbitrary orthogonal matrix in $\mathfrak{B}^{\mathcal{C}}_{M}( \mat{H}_{\circ})$ that is also orthogonal to $\mat{I}_K$, i.e. $\tr\{\mat{B}_1\}=\tr\{\mat{B}_1^\mathrm{T} \mat{I}_K\}=0$. Having chosen the first elements up to $\mat{B}_l$, take $\mat{B}_{l+1}$ (if it exists) as an orthogonal matrix in $\mathfrak{B}^{\mathcal{C}}_M( \mat{H}_{\circ})$ which is also orthogonal to all the previous elements, i.e. $\tr\{\mat{B}_{l+1}^\mathrm{T} \mat{B}_i\}=0$, for every $i\leq l$. We claim that the family $\{\mat{B}_i\}_{i=1}^{d-1}$ is Hurwitz-Radon, where $d$ is the dimension of $\mathfrak{B}^{\mathcal{C}}_M( \mat{H}_{\circ})$.\\
Indeed for each $i\neq j$, $\mat{B}_i+\mat{B}_j$ is contained in $\mathfrak{B}^{\mathcal{C}}_{M}( \mat{H}_{\circ})$ and hence is orthogonal up to a multiplicative constant. As $\mat{B}_i$ and $\mat{B}_j$ are also orthogonal, this leads to $\mat{B}_i^\mathrm{T}\mat{B}_j+\mat{B}_j^\mathrm{T}\mat{B}_i=\beta \mat{I}_K$ for some real number $\beta$. This along with the assumption $\tr\{\mat{B}_i^\mathrm{T}\mat{B}_j\}=0$ results in $\mat{B}_i^\mathrm{T}\mat{B}_j+\mat{B}_j^\mathrm{T}\mat{B}_i=\mat{0}$. Letting $j=0$, we also get $\mat{B}_i^\mathrm{T}+\mat{B}_i=\mat{0}$ for every $i\neq 0$. The proof for $\mat{\mathscr{B}}^{\mathcal{C}}_*$ is identical.\\
For the second part, we again note that $\mat{B}$ can be written as $\alpha\mat{I}+\mat{B}'$ where $\mat{B}'$ is a skew-symmetric matrix. As the matrix $\mat{B}'$ is skew-symmetric, there exist \cite{horn} a real orthogonal matrix $\mat{Q}$ and a family $\{\mat{\Sigma}_i\}_{i=1}^{K'}$ of matrices of the form
$\begin{bmatrix}
0 & \beta_i\\
-\beta_i & 0
\end{bmatrix}$
for $\beta_i\in \mathbb{R}$, such that
$\mat{B}'=\mat{Q}\mat{\Sigma}\mat{Q}^{\mathrm{T}}$, where $\mat{\Sigma}$ is the block-diagonal matrix defined as $\diag(\mat{\Sigma}_1, \cdots, \mat{\Sigma}_{K'})$, i.e., the matrix that contains $\mat{\Sigma}_i$'s on its diagonal, and zero entries every where else. Using $\mat{Q}\,\mat{Q}^{\mathrm{T}}=\mat{I}$, we obtain $\det(\alpha\mat{I}+\mat{B}')=\det(\alpha\mat{I}+\mat{\Sigma})$, hence the positivity of the determinant is clear.
\end{IEEEproof}

\begin{remark}
A byproduct of this theorem is that when $K$ is odd, the spaces $\mathfrak{B}^{\mathcal{C}}_{M}( \mat{H}_{\circ})$ and $\mat{\mathscr{B}}^{\mathcal{C}}_*$ are always one dimensional and are generated by the identity matrix. This has already been proved in \cite{BI}.
\end{remark}

\begin{example}\label{Example}
For the Alamouti code, we have $\mat{C}_1=\mat{I}_2$, $\mat{C}_2=i \mat{\Omega}_2$,
$C_3=\begin{bmatrix} 0 & 1\\ -1 & 0 \end{bmatrix}$, and $C_4=i \mat{\Omega}_4$, where $\mat{\Omega}_2:=\begin{bmatrix} 1 & 0\\ 0& -1 \end{bmatrix}$, and
$\mat{\Omega}_4:= \begin{bmatrix} 0 & 1\\ 1& 0 \end{bmatrix}$.
One can easily show that $\mat{\mathscr{B}}^{\mathcal{C}}_*$ is the $4$-dimensional space generated by $\{\mat{I}_4, \mat{C}_3\otimes \mat{I}_2, \mat{\Omega}_4\otimes \mat{C}_3, \mat{\Omega}_2\otimes \mat{C}_3\}$.
\end{example}

\section{Final remarks}

Let us go back to the effect of the channel ambiguity on the estimation of the input-data vector. We saw in Equation \eqref{rotated signal at the receiver} that the effect of the channel ambiguity is an orthogonal transformation. But with the results we developed in the last section we know a lot more about this orthogonal ambiguity. Indeed, let us assume that we have found a solution $\hat{\mat{h}}$ to the channel estimation equation, i.e., Equation \eqref{estimation of the channel vector}. Using this solution we get an estimation for the input signal which we denote by $\hat{\mat{s}}$. By Equation \eqref{decoding the data signal} or \eqref{rotated signal at the receiver}, we know that in the absence of noise, the true input signal $\mat{s}$ is related to the estimated $\hat{\mat{s}}$ as follows
$$
\mat{s}=\frac{\|\tilde{\mat{h}}\|}{\|\mat{h}_{\circ}\|}
\mat{B}_{\hat{\mat{h}}} \,\tilde{\mat{s}}.
$$
We know that $\mat{B}_{\hat{\mat{h}}}$ lies in $\mathfrak{B}^{\mathcal{C}}_{M}( \mat{H}_{\circ})$. As this space depends on the channel realization, we can not compute it a priori. But if $M\geq M_*^{\mathcal{C}}$,
then $\mathfrak{B}^{\mathcal{C}}_{M}( \mat{H}_{\circ})$ is equal to $\mat{\mathscr{B}}^{\mathcal{C}}_*$ which is a deterministic space, independent of the channel matrix or the number of receive antennas, and more important, it can be explicitly computed for every OSTB. In this case the true input-data signal $\mat{s}$ lies in the following space
$$
\mathcal{S}^{\mathcal{C}}(\tilde{\mat{s}}):=\{\mat{B} \,\tilde{\mat{s}}\;\big{|}\; \mat{B}\in \mat{\mathscr{B}}^{\mathcal{C}}_*\}.
$$

This characterisation may be used to design multi-dimensional constellation schemes for the input-data vectors $\{\mat{s}_i\}_i$ such that they all stay invariant under the elements of the space $\mat{\mathscr{B}}^{\mathcal{C}}_*$, which as we showed, are pure rotations. In particular, when the dimension of $\mat{\mathscr{B}}^{\mathcal{C}}_*$ is small compared to $K$, this approach might generate high-rate coding schemes that are still identifiable. We leave this subject for a future project.

\appendix
\section{Proof of Theorem \ref{thm1}}\label{appen}

\begin{lem}\label{lemthm}
Let $\mat{D}$ be a real diagonal matrix of size $m\times m$ with diagonal elements $\lambda_1 \geq \lambda_2 \geq \cdots \geq \lambda_m$, and $q$ be a
positive integer not larger than $m$. Let ${q^-}$ be the largest index such that $\lambda_{{q^-}}>\lambda_q$, and ${q^+}$ be the largest index for which $\lambda_{{q^+}}=\lambda_q$.\\
Then a matrix $\mat{Q}\in \mathscr{U}^{m\times q}$ is in $\mathcal{A}_q^{\mat{D}}=\argmax_{\mat{Q} \in \mathscr{U}^{m\times q}}\tr\bigl\{\mat{Q}^{\mathrm{T}}\mat{D}\mat{Q}\bigr\}$ if and only if
its column space equals
\begin{equation}\label{e6.1.3}
\langle\mat{e}_1, \mat{e}_2, \dots ,\mat{e}_{{q^-}} \rangle \oplus \mathcal{W}_{q-{q^-}}\,,
\end{equation}
where $\mat{e_i}$'s are the elementary bases of $\mathbb{R}^m$, $\langle\mat{e}_1, \mat{e}_2, \dots ,\mat{e}_{{q^-}} \rangle$ is the subspace generated by the mentioned vectors, and $\mathcal{W}_{q-{q^-}}$ is some ($q-{q^-}$)-dimensional subspace of $\langle\mat{e}_{{q^-}+1}, \mat{e}_{{q^-}+2}, \dots ,\mat{e}_{{q^+}} \rangle$.
\end{lem}

\begin{IEEEproof}
Let's denote by $\mat{x}_i$ the rows of $\mat{Q}$, i.e.
$$
\mat{Q}^\mathrm{T}=
\begin{bmatrix}
\mat{x}_1^\mathrm{T} & \mat{x}_2^\mathrm{T}& \cdots & \mat{x}_m^\mathrm{T}
\end{bmatrix}.
$$
Then the following equations can be easily verified:
\begin{equation}\label{value of the trace}
\tr \{ \mat{Q}^{\mathrm{T}} \mat{D} \mat{Q}\}= \sum_{i=1}^m \lambda_i \| \mat{x}_i \|^2,
\end{equation}
and
\begin{equation}\label{e6.1.2}
q=\tr \{ \mat{I}_q \} = \tr \{ \mat{Q}^{\mathrm{T}} \mat{Q}\}= \sum_{i=1}^m \| \mat{x}_i \|^2.
\end{equation}
Now by augmenting $m-q$ orthonormal columns to $\mat{Q}$, we may complete it into an orthogonal matrix of size $m \times m$. It is evident that each row of the augmented matrix should have a unit norm. So the norm of each row of $\mat{Q}$ is at most $1$. In other words:
\begin{equation}\label{e6.1.4}
\forall i=1, \dotsc , m \quad : \quad  \| \mat{x}_i \|^2 \leq 1.
\end{equation}
So by Equations \eqref{value of the trace}, \eqref{e6.1.2} and \eqref{e6.1.4}, we easily get
$$
\begin{aligned}
\tr \{ \mat{Q}^{\mathrm{T}} \mat{D} \mat{Q}\}
                                   & \leq \sum_{i=1}^{{q^-}} \lambda_i + \sum_{i=1}^{{q^-}} \left \{ (\lambda_{q}-\lambda_{i}) (1-\| \mat{x}_i \|^2)  \right \}+\\
                                   &   +\lambda_q (q-{q^-}) +(\lambda_{{q^+}+1}-\lambda_q)\bigl \{q -\sum_{i=1}^{{q^+}} \| \mat{x}_i \|^2\bigr \}.
\end{aligned}
$$
It is also evident that the right-hand-side expression of this inequality is less than or equal to $\sum_{i=1}^{{q^-}} \lambda_i +\lambda_q (q-{q^-})$, with equality \underline{if and only if}
\begin{equation}\label{e6.2}
\forall \; i=1, \dotsc , q^- \quad : \quad  \| \mat{x}_i \|^2 = 1\,,
\end{equation}
and
\begin{equation}\label{e6.2.2}
\sum_{i={q^-}+1}^{{q^+}} \| \mat{x}_i \|^2=q-{q^-}\,.
\end{equation}
So $\sum_{i=1}^{{q^-}} \lambda_i +\lambda_q (q-{q^-})$ is an upper bound on $\tr \{ \mat{Q}^{\mathrm{T}} \mat{D} \mat{Q}\}$ which is independent of $\mat{Q}$. It is also evident that a necessary condition for a matrix $\mat{Q}\in\mathscr{U}^{m\times q}$ to achieve this upper bound, is to satisfy \eqref{e6.2} and \eqref{e6.2.2}.
On the other hand, one can be easily verify that any matrix $\mat{Q}\in\mathscr{U}^{m\times q}$ satisfying \eqref{e6.2} and \eqref{e6.2.2}, also achieves this upper bound. So a matrix $\mat{Q}$ in $\mathscr{U}^{m\times q}$ achieves this upper bound if and only if it satisfies conditions \eqref{e6.2} and \eqref{e6.2.2}. Now we will verify that a matrix $\mat{Q}$ in $\mathscr{U}^{m\times q}$ satisfies conditions \eqref{e6.2} and \eqref{e6.2.2} if and only if there exist a matrix $\mat{W}$ of size $({q^+}-{q^-})\times(q-{q^-})$ with orthonormal columns and an orthogonal matrix $\mat{B}$ such that
\begin{equation}\label{e6.2.3}
\mat{Q}=
\begin{bmatrix}
%\begin{matrix}
\mat{I}_{{q^-}} & \mat{0}_{q\times(q-{q^-})}\\
\mat{0}_{({q^+}-{q^-})\times {q^-}} & \mat{W}\\
%\end{matrix}\\
\mat{0}_{(m-{q^+})\times q^-} & \mat{0}_{(m-{q^+})\times (q-q^-)}
\end{bmatrix} \mat{B}.
\end{equation}
Suppose that a matrix $\mat{Q}$ in $\mathscr{U}^{m\times q}$ satisfies conditions \eqref{e6.2} and \eqref{e6.2.2}. By \eqref{e6.1.2}, all the last $m-{q^+}$ rows of $\mat{Q}$ have to be zero. Furthermore, the first ${q^-}$ rows of $\mat{Q}$ must be orthonormal. To verify this, we eliminate the last $m-{q^+}$ rows of $\mat{Q}$ to obtain the sub-matrix $\mat{Q'}$. It is evident that  $\mat{Q'}$ inherits from $\mat{Q}$ its property of having orthonormal columns. Now by augmenting ${q^+}-q$ extra orthonormal columns to $\mat{Q'}$ we may complete it to an orthogonal matrix $\mat{Q''}$ in $\mathbb{R}^{{q^+}\times {q^+}}$. Clearly the rows of $\mat{Q''}$ have to be orthonormal. In particular, its first ${q^-}$ rows must have unit norm. But the first ${q^-}$ rows of $\mat{Q}$ have already unit norm. So in its first ${q^-}$ rows, $\mat{Q''}$ differs $\mat{Q}$ only by having extra zero entries. This shows that the first ${q^-}$ rows of $\mat{Q}$ are indeed orthonormal. Now we can construct matrix $\mat{B}$ by letting its first ${q^-}$ rows be the first ${q^-}$ rows of $\mat{Q}$, and its last $q-{q^-}$ rows, be given by the Gram-Schmidt procedure such that the rows of $\mat{B}$ form an orthonormal basis for the row space of $\mat{Q}$. This construction is possible because $\mat{Q}$ is in $\mathscr{U}^{m\times q}$ and hence of rank $q$.

Finally, for any matrix $\mat{Q}$ in $\mathscr{U}^{m\times q}$, Equality \eqref{e6.2.3} is equivalent to $\mat{Q}$ having its column space equal to the space given by Equation \eqref{e6.1.3}.
\end{IEEEproof}

\begin{IEEEproof}[Proof of Theorem \ref{thm1}]
We take the following notation:\\
Let $\lambda_1$, $\lambda_2$, \dots, $\lambda_m$ list all the eigenvalues of matrix $\mat{P}$ in decreasing order, and $\mat{v}_1$, $\mat{v}_2$, \dots, $\mat{v}_m$ be their corresponding orthonormal eigenvectors. Let also ${q^-}$ be the largest integer such that $\lambda_{{q^-}}<\lambda_q$ and ${q^+}$ be the largest integer for which $\lambda_{{q^+}}=\lambda_q$.\\
We should prove that a matrix $\mat{Q}\in \mathscr{U}^{m\times q}$ is in $\mathcal{A}_q^{\mat{P}}=
\argmax_{\mat{Q} \in \mathscr{U}^{m\times q}}\tr\bigl\{\mat{Q}^{\mathrm{T}}\mat{P}\mat{Q}\bigr\}$ if and only if
its column space equals
\begin{equation}\label{e6.3.1}
\langle\mat{v}_1, \mat{v}_2, \dots ,\mat{v}_{{q^-}} \rangle \oplus \mathcal{W}_{q-{q^-}},
\end{equation}
where $\mathcal{W}_{q-{q^-}}$ is an arbitrary ($q-{q^-}$)-dimensional subspace of $\langle\mat{v}_{{q^-}+1}, \mat{v}_{{q^-}+2}, \dots ,\mat{v}_{{q^+}} \rangle$.
$\mat{P}$ can be represented as
$%\label{e6.3}
\mat{P}=\mat{U}\mat{D}\mat{U}^{\mathrm{T}}
$, where $\mat{D}$ is a diagonal matrix having $\lambda_1$, $\lambda_2$, \dots, $\lambda_m$ as its diagonal elements and $\mat{U}$ is the orthogonal matrix having $\mat{v}_1$, $\mat{v}_2$, \dots, $\mat{v}_m$ as its columns.

Clearly, for any $\mat{Q}$ in $\mathscr{U}^{m\times q}$, we have $\mat{Q}\in \mathcal{A}_q^{\mat{P}}$ if and only if $\mat{U}^{\mathrm{T}}\mat{Q}\in \mathcal{A}_q^{\mat{D}}$. So by
Lemma \ref{lemthm}, $\mat{Q}\in \mathcal{A}_q^{\mat{P}}$ if and only if
there exists a matrix $\mat{W}$ of size $({q^+}-{q^-})\times(q-{q^-})$ with orthonormal columns and an orthogonal matrix $\mat{B}$ such that
$$
\mat{U}^{\mathrm{T}}\mat{Q}=
\begin{bmatrix}
%\begin{matrix}
\mat{I}_{{q^-}} & \mat{0}_{q\times(q-{q^-})}\\
\mat{0}_{({q^+}-{q^-})\times {q^-}} & \mat{W}\\
%\end{matrix}\\
%\mat{0}_{(m-{q^+})\times q}
\mat{0}_{(m-{q^+})\times q^-} & \mat{0}_{(m-{q^+})\times (q-q^-)}
\end{bmatrix} \mat{B}.
$$
Finally, multiplying this equation from the left by $\mat{U}$ completes the proof.
\end{IEEEproof}

\bibliographystyle{plain}
%\bibliography{MIMO-Bibliography}

\end{document}